# Helping IT and OT Defenders Collaborate

Glenn A. Fink
National Security Directorate
Pacific Northwest National Laboratory
Richland, Washington, USA
Glenn.Fink@pnnl.gov

Penny McKenzie
National Security Directorate
Pacific Northwest National Laboratory
Richland, Washington, USA
penny.mckenzie@pnnl.gov

*Abstract*— **Cyber-physical systems, especially in critical infrastructures, have become primary hacking targets in international conflicts and diplomacy. However, cyber-physical systems present unique challenges to defenders, starting with an inability to communicate. This paper outlines the results of our interviews with information technology (IT) defenders and operational technology (OT) operators and seeks to address lessons learned from them in the structure of our notional solutions. We present two problems in this paper: (1) the difficulty of coordinating detection and response between defenders who work on the cyber/IT and physical/OT sides of cyber-physical infrastructures, and (2) the difficulty of estimating the safety state of a cyber-physical system while an intrusion is underway but before damage can be effected by the attacker. To meet these challenges, we propose two solutions: (1) a visualization that will enable communication between IT defenders and OT operators, and (2) a machine-learning approach that will estimate the distance from normal the physical system is operating and send information to the visualization.**

## I. INTRODUCTION

For years the NSA and FBI have been publicly warning utility owners that cyber attacks on their physical infrastructures were imminent [1]. In December of 2015 and 2016 the world saw the first major manifestation of the cyber vulnerability of infrastructural cyber-physical systems as Ukraine's electric grid was repeatedly attacked, resulting in a quarter million Ukrainians going without power in winter temperatures [2]. More recently, the VPNFilter malware was noted to have the capability to monitor devices on control-systems networks [3]. Networked computer systems have become an attack vector for cyber-physical infrastructures that can produce severe physical and political effects. However, meeting this threat requires coordinating defenses between cyber security staff and physical plant operators. Human effectiveness is a challenge for cyber-physical systems (CPS) security, but human influence adds both resilience and complexity to cyber-physical systems. Research is required to enable the needed collaboration.

We present two problems in this paper and two matching solutions. The first problem is the difficulty of coordinating detection and response between defenders who work on the cyber (Information Technology, IT) and physical (Operational Technology, OT) sides of infrastructures. The proposed matching solution is a visualization that will enable communication between IT defenders and OT operators. The second problem is the difficulty of estimating the safety state of a cyber-physical system while an intrusion is underway but before the attacker can do damage. The proposed solution is a machine-learning approach that will estimate the distance from normal the physical system is operating and send information to the visualization. This work outlines the results of our interviews with IT defenders and OT operators and seeks to address lessons learned from them in the structure of our notional solutions.

## II. BACKGROUND

The United States President's National Security Telecommunications Advisory Committee (NSTAC) defines CPS as follows:

> Decentralized network of objects (or devices), applications, and services that can sense, log, interpret, communicate, process, and act on a variety of information or control devices in the physical environment. These devices range from small sensors on consumer devices to sophisticated computers in industrial control systems (ICS). Ultimately, the devices have some type of kinetic impact on the physical world, whether directly or through a mechanical device to which they are connected [4].

Protecting OT/CPS in critical infrastructures is a principle concern for any nation. Cyber weapons may be as lethal to infrastructures as kinetic warfare, and, as Stuxnet demonstrated, not even air-gapped systems are safe [5]. Software-borne weapons in embedded systems may stay dormant or subtly active for months or years as the remote attackers wait for the right moment to attack. Detection of attacks from the physical side alone may be very difficult, and once attackers gain access to a control system they may cause a variety of distracting problems. There are five stages the attacker goes through when hacking a complex cyber-physical control system [6]:

1. Access: gained through traditional hacking approaches
2. Discovery: determining the unique physical and logical functions of the victim system
3. Control: learning the limits of control available via software
4. Damage: deciding on and producing the kinds of damage he or she would like to inflict, and finally
5. Cleanup: causing the operators to believe what the attacker wants them to believe about the attack.

The access phase presents the best hope for discovering attackers in cyber-physical systems from the cyber side. From the physical side, noting anomalies during the discovery and control phases is the most likely way to stop an attack before damage occurs. The discovery and control phases may require months or years of development after initial access is gained before noticeable damage may be inflicted. Thus, unless attackers are caught during the discovery and control phases, correlating the attack activities back to the cyber indicators of the access phase may be very difficult for the defenders.

Detection of attacks from the cyber side is difficult because, typically, cyber operators do not understand the protocols, physical processes, or potential consequences of attacks on the physical system. Even when defenders detect attackers gaining access, it is difficult to understand the ultimate goal of attacks or to determine whether the attackers have achieved their purposes. Similarly, the cyber side has complexities not well understood by the control-systems operators. Both groups of operators need to understand previous cyber attacks so they can understand potential risks to the physical plant in the future.

*A. Welcome to XYZ, Corp.*

For our study, we refer to network configurations and security conditions at a fictitious company, XYZ Corp., that is made of a conglomeration of real companies from whose systems, networks, and operators we gathered information. Fig. 1 shows the basic network setup of XYZ Corp.

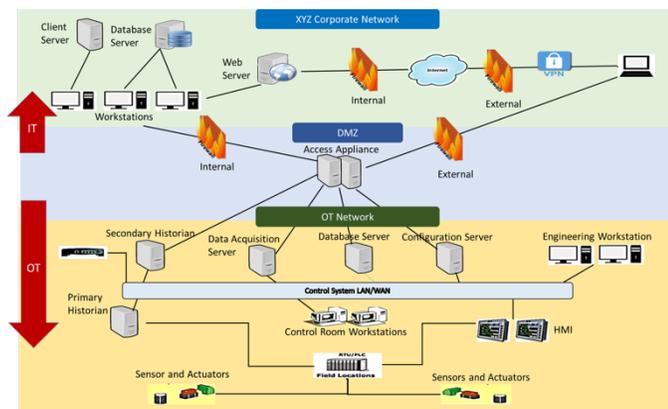

Fig. 1. XYZ Corporation IT/OT Network.

As shown in the figure, XYZ's network is divided into a corporate Local Area Network (LAN), a demilitarized zone (DMZ), and an ICS network (which XYZ Corp. calls the *facilities network*). Each zone is cordoned off from the others and from the internet via a firewall and access-control rules. Controlled remote access allows the site to collect data from rented facilities that have their own networks and allows collaborators authenticated access to data and OT systems.

Ideally, there would be only a single, tightly controlled connection between the corporate LAN and the control systems LAN. However, there may be many access points (some of which cross air gaps) implying inconsistent access controls and multiple paths for an attacker to target. Systems on the corporate LAN need occasional access to systems on the facilities network for control/status and to perform updates. Even air gaps must be considered high-latency internet connections.

At XYZ, the cyber-physical systems in the facilities network are generally composed of remote telemetry units (RTUs), programmable logic controllers (PLCs), intelligent electronic devices (IEDs), and the sensors and actuators that control and read from the physical devices. We collectively refer to these as control systems devices or controllers. There is also a set of IT servers that collect and provide access to the data from the controllers, and human-machine interface (HMI) devices that enable a command and status channel to the human operators.

As control systems devices increase in computational and networking capabilities, they become more susceptible to cyber attack. They are designed for long life without updates. Unfortunately, this means that once an exploit is discovered for a controller, it may be vulnerable for life. These components are not designed to manage their own security and they simply accept and run any command or firmware update they receive. Thus, attackers can surreptitiously re-image devices. Discovering this duplicity and repairing the damage it causes can be an extremely complex matter.

Cyber-physical systems at XYZ include HVAC and building control and automation for lab spaces, air flow handlers, scientific instruments, and more. There are an estimated 20,000 addressable networked devices at XYZ Corp. involved in cyber-physical systems, but the breakdown into classes is unknown. Perhaps one-quarter (5,000 addresses) belong to actual physical devices and to associated research equipment. Several of XYZ's newest buildings publish sensor information to the corporate network as to help bring operational CPS data to researchers.

Every control system is a unique combination of sensors, actuators, controllers, and computers connected by excessively many semi-interoperable proprietary protocols. OT staff specialize in knowing the idiosyncrasies of these highly diverse systems and their protocols. Today's OT systems are process-based; they do not do detection based on the information content of the network traffic, and there is little commercial impetus to do so. Frequent errors are generally tolerated if they do not produce downtime. Error that do not produce downtime are seldom investigated, and no logs of commands sent to devices are kept. None of these systems are deemed critical from an IT perspective, so logging seems to be an unnecessary expense.

Little or no security is included in OT protocols. For instance, Modbus protocol simply provides a way to write to and read from devices without any notion of authentication, confidentiality, integrity, or authorization. Encryption and authentication protocols require costly key management processes, and it is more cost effective to simply cordon these systems off on a secured network. This is efficiency, not bad design. But newer devices on facilities networks may have uncontrolled cellular connectivity to the Internet. When control systems are unintentionally exposed to the broader threat-space of the internet, problems may arise.

Generally, OT devices that have serial-only connections are grouped into a "field bus" network under a Network Area Controller (NAC) device connected to an Ethernet network via TCP/IP. Native-protocol commands for the devices on the field bus are tunneled though TCP/IP and passed on to the proper device by the NAC. Typically, no syntactic or semantic validation or stateful inspection of the tunneled commands is done because this would require expensive emulation of all the types of devices on the NAC. At XYZ, the NAC's primary security control is its isolation of the facilities network. The NAC uses a firewall that logs and controls its access from the XYZ network.

Some cyber-physical systems use Windows-based controllers, and many of these systems at XYZ Corp. have Windows Logging Service (WLS) installed. WLS takes information from the native Microsoft Windows log events, correlates security and resource usage information to it, and repackages it in XML in a key/value format like what the Unix syslog facility provides. WLS data would provide the most comprehensive set of host-based data for the traditional IT systems found within OT networks. However, the actual controllers (PLCs/RTUs/etc.) do not run Windows. Instead they may run some completely custom application without an operating system as is done on Application-Specific Integrated Circuits (ASICs). For them, security software is not an option.

Some XYZ buildings are leased facilities where the cyber-physical building systems are on a private network, owned and secured by the building owner. Physical and cyber security controls in leased buildings on campus are not mandated by XYZ and may be less rigorous. No cyber intrusions have yet interfered with cyber-physical systems in the leased building's network. However, there have been reports of research systems attached to one leased building's network being exposed to cyber attack. Although these buildings may control HVAC for large XYZ operational computing systems, because XYZ does not manage their networks, it considers them untrusted facilities and does not allow any physical connection between the leased buildings' networks and the XYZ facilities network.

### III. Approach to a solution

We interviewed OT operations personnel, cyber defenders, and researchers who work with cyber-physical systems to understand the security needs in this space. We sought to understand whether cyber incidents had ever been known to cause problems in a cyber-physical system and the proximate causes of any such incidents. In this section, we will examine the findings of our interviews from three perspectives: the owner/operators of physical systems that have become cyber-physical systems, the cyber defenders, and the researchers.

#### A. Physical Systems (OT) Perspective

We interviewed building managers, physical plant operators, and control-systems engineers to understand their perspective on how to improve the security of cyber-physical systems. For most of the half century IT has existed, IT and OT groups have not had much reason to interact. Frequently, IT groups are unfamiliar with the function of OT systems and have little interest in them. From the OT side, as long as there was no credible threat from the network, OT staff saw no reason to open regular dialogue with cyber defenders. Nobody on the OT side watches the operational cyber side on a day-to-day basis. OT staff are frequently unaware of the tools techniques and procedures used to secure the IT networks. OT staff are not part of the discussion when cyber security policy is changed and little thought is given to understanding how changes affects the cyber-physical systems in control systems networks. This means that policies, updates, and procedures that could be of security benefit may not be applied to the standard IT systems (like Windows servers) operating within the OT facilities network. Although both sides believe dialogue would be beneficial, no formal arrangement exists at XYZ Corp. yet.

When OT staff or researchers at XYZ add a new TCP/IP capable device, IT cyber security personnel must be notified. But otherwise, each team is a black box to the other. As IT and OT systems have converged, and many cyber-physical devices became web-based, there is growing interest and pressure to collaborate.

The OT staff would find it useful to know what cyber threats are being aimed at their systems. They want to understand what kind of attacks have happened in the past so they can learn lessons and understand potential risks. The expectation is that if an anomalous cyber event happens directed at OT systems, the IT cybersecurity staff would notify the OT staff. The cybersecurity staff at XYZ do not yet regularly provide threat information to OT staff, but they do monitor the average traffic destined for each cyber-physical device. They send a list of IP addresses and amounts of traffic to OT operators weekly. OT staff review this list to see if there are dramatic (>10% to 20%) changes. Most of the anomalies they discover in this way arise from researchers making changes to their data-collection devices residing in the OT network. Occasionally, a misconfigured device is discovered in this way.

OT staff we interviewed related several classes of threat related to cyber attack and outlined their potential consequences. This information is shown in TABLE I.

TABLE I. THREATS OF CONCERN TO XYZ'S OT STAFF

| Threat | Consequence |
| --- | --- |
| Unauthorized access | Loss of accountability and control; down time; equipment damage |
| Loss or manipulation of data | Intellectual property loss, scientific progress hampered, system malfunction |
| Laboratory vent hoods or safety systems taken offline by an attack | Injury or loss of life; deleterious physical effects; release of potentially harmful or deadly gasses |
| Server room chillers degraded or offline | Computer systems could sustain significant damage |
| Negative pressure facilities compromised | Chemical, biological, and radioactive contamination of environment and staff |
| Overpressurization of water lines | Physical damage; monetary loss |
| Uninterruptible power supply compromise | Inability to bring computer systems down in an orderly fashion in an emergency |
| Alteration of set points in critical HVAC systems | Damage, degradation, or loss of computer systems |

#### B. Cyber Security Perspective

We interviewed several cyber security professionals at XYZ who are familiar with cyber-physical systems and the unique defensive challenges they represent. We noted that cyber security staff also lament the lack of communication with OT staff. There is no organizationally prioritized channel of formal communication between the IT and OT teams. Problems on the

OT side are seldom if ever reported to cyber staff. Cyber security personnel pass warnings on to OT personnel but they suspect OT staff may not understand or act upon the warnings effectively. Cyber analysts believe that OT staff who find malware on their systems will simply remove the malware files and reboot the systems without considering how to prevent future infections. Many OT systems have no possibility of host-based protection.

Cyber analysts know which IP address ranges generally belong to cyber-physical systems, but only the OT staff understand the protocols, capabilities, and proper function of control systems. Cyber analysts often do not understand the underlying physical system well enough to act or notify the system owners of something suspicious. Cyber analysts frequently cannot even be certain whether these systems need to be patched because active scanning frequently causes failures. An infamous incident at XYZ Corp. happened when cyber assessors tried to scan an IT network that, unknown to them, was connected to an OT network. When they started a Nessus scan, it activated sprinklers in a server room causing millions of dollars in damage.

Lack of visibility into the proper workings of cyber-physical systems is the gravest concern of cyber analysts. The threat and business risk are not even roughly quantified for what would happen if a given CPS failed in specific ways. Existing security plans instead focus on documenting connectivity and quantifying risk from a purely IT perspective.

Although most OT systems are listed as noncritical, most cyber staff recognize that failures in the physical plant could be exceptionally costly. Although this would tend to make one consider these systems high-value assets, cyber security staff are already tasked with more priorities than can be addressed. Listing them as critical would not help unless more resources were available to handle the workload [7]. Thus, these systems remain listed as non-critical.

*C. Research Perspective*

PNNL has developed several tools for monitoring cyber-physical infrastructure to improve performance and security. Among them, we cover two that will serve as data sources for our proposed solution: VOLTTRON™ and SerialTap.

*1) VOLTTRON™*

VOLTTRON™ is an open-source, secure, extensible, and modular building-management technology that enables mobile and stationary software agents to perform information gathering, processing, and control actions. VOLTTRON can independently manage a wide range of applications, such as HVAC systems, electric vehicles, distributed energy or whole-building loads, leading to improved operational efficiency and energy and cost savings. Our researchers have gained permission to place VOLTTRON data-collection agents on XYZ's facilities network and on the networks of its leased buildings and remote facilities (Fig. 2). Each instance forwards data to the XYZ corporate network via a special access agent located in the XYZ network DMZ. There are 12 VOLTTRON servers running in each of 12 buildings. Instances of VOLTTRON communicate via VOLTTRON Information Protocol (VIP), a message passing protocol based on ZeroMQ (http://zeromq.org). Building systems data from VOLTTRON agents is made available to researchers for energy-related objectives and also for advanced cyber detection research in cyber-physical systems.

No outbound access is available directly from sensors. Instead they forward data to the "Management Central" VOLTTRON instance in the XYZ network. Commands may be sent to the building management system (BMS) on the facilities network from authorized devices only. The BMS commands these OT systems via field bus protocols tunneled under IP. The 12 VOLTTRON systems on the facilities network query the BMS to obtain their information. These systems relay status information to the VOLTTRON Management Central system. XYZ OT data is made available to certain external collaborators via the VOLTTRON Passthrough instance in the XYZ DMZ. Other access to this information is available via a web portal that passes through a traffic scanner for security.

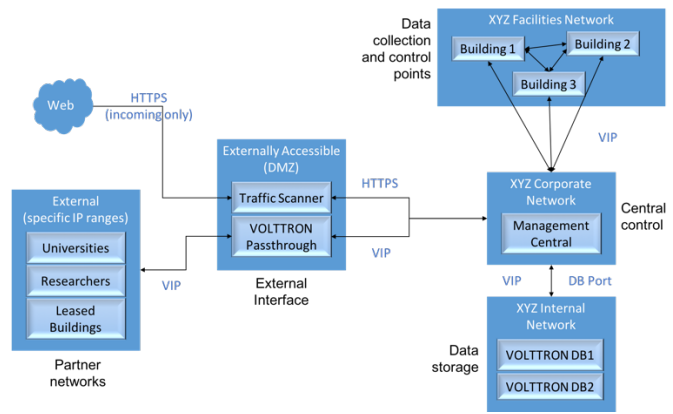

Fig. 2. Architecture of building-controls sensors at XYZ.

Although there has never been an OT fault that was known to be correlated with a cyber incident, it is quite possible to self-induce a denial of service by doing too many simultaneous queries on the VOLTTRON instances because these in-turn query the sensors causing them to do extra work. This happened once when researchers were setting up a demonstration simultaneously showing all the information that can be gained from the many VOLTTRON sensors. If attackers could subvert a VOLTTRON node they could gain read-only access to see field bus protocol commands. While this may facilitate an attacker's reconnaissance during the discovery and control phases, VOLTTRON would be much harder to subvert than most components on the facilities network because it is security hardened.

*2) SerialTap*

SerialTap is another PNNL tool that enables enterprise cyber security tools to monitor communications on serial field busses by tapping into the serial bus and tunneling the traffic it finds there via Ethernet packets to the enterprise LAN for analysis by cyber analysts and their tools. The visibility VOLTTRON provides to OT TCP/IP networks SerialTap affords for serial networks with physical media like RS-232 cabling. SerialTap is designed to fail closed to avoid obstructing communications if the SerialTap device fails. SerialTap is very inexpensive so that it could be placed on multiple network segments economically.

The cost of segment-wise monitoring depends on the topology of the field bus. In a bus topology, a single monitor anywhere on the bus can see the traffic from the entire segment because all traffic is broadcasted to every node. In contrast, star topologies have only point-to-point serial connections between the PLC/RTU and each controlled/monitored device, so monitors must be placed inline as pass-throughs on each spoke of the network. Ring topologies are also sometimes used, and, depending on the protocols used, every station on the ring may still see every command. But if each device consumes the commands sent to it, pass-through monitors must be placed at every network segment to capture traffic. The more monitors required, the greater the expense of implementation.

SerialTap and VOLTTRON can provide important sources of OT data to the cyber analyst and enable better situational awareness decision support. Additionally, they will expose hidden attacks and allow preservation of transient data that may contain clues about an intruder's purpose and methods.

### D. Lessons learned from the interviews

From our interviews of staff on both sides it seems each group has competence in its own area but fears that the other group does not have sufficient competence in that area to help. On the physical side staff care about availability of services without concern for failure analysis. From the cyber security side, root-cause analysis is critical to prevent reoccurrence of the same failure.

A major problem in XYZ Corp. and elsewhere is the lack of communication between IT/cyber and OT/physical subject matter experts. In most organizations, it would be unusual for the IT and OT groups to meet regularly. The two groups have operational specialties and vocabularies that differ significantly, and communication is often *ad hoc*. OT devices are complex, not well understood, and insufficiently monitored for security. Ideally IT and OT staff would always be aware of the issues on OT networks: cyber threats, performance benchmarks, noisy or malfunctioning devices, etc. OT staff should communicate these issues back to IT, and IT staff must learn to communicate clearly to the OT operators.

While analysts typically want to have all the data, all cyber defenders really need to know about OT systems is the degree of anomaly they are experiencing. This would require collection of data on the OT networks using something like SerialTap/VOLTTRON and requires automatic data analysis to determine quickly the level of anomaly and potential threat. Employing these technologies on the OT networks allows operators to query the status of the system and intervene directly where necessary, bringing nearly as much visibility to the OT side as is available to cyber defenders on the IT side.

The gap identified by our interviews outlines the shape of a communications tool to enable both cyber security analysts and control systems operators to react in near real time to cyber attack. Additionally, the tool would act as a general communication solution between the two groups. The reason such a tool does not yet exist is that both cyber and OT data are difficult to gather in one place for analysis and analysis.

## IV. PROPOSED SOLUTION

The problems to be solved are lack of collaboration between IT and OT staff and difficulty understanding the state of the cyber-physical infrastructure as a whole. To solve the first problem, we propose a visual interface that could be used equally well by both groups as a conversation starter. This visualization would be driven by the solution to the second problem: a data-driven deep-learning model that ingests cyber-physical data from the enterprise and correlates performance impact to prior suspicious cyber activity. The proposed model will learn a representation of normal behavior, evaluate variations and excursions from automatically discovered normal operating bounds, and present them using the visualization. The technology is designed to recognize distributed threats across the interwoven layers of OT/IT architecture over long time periods and provide measurable improvement to current state-of-the-art technologies for OT/IT intrusion detection.

### A. Visualization Prototype

The visualization prototype is designed to present the degree of anomaly in OT systems in a way that is easily understandable to both cyber analysts and to control systems operators. The visualization will help human analysts to communicate and to locate the most anomalous parts of the system to prioritize investigate at a high level. The cyber analyst only needs to know how anomalous the cyber-physical system is, not what exactly is wrong with it. For the control systems specialist, the level of anomaly shown will indicate that further investigation is needed. The resulting trend visualization concept (Fig. 3) reduces the anomaly measure of an entire system to a set of bars that each represent a dimension of abnormality and whether it is increasing or decreasing.

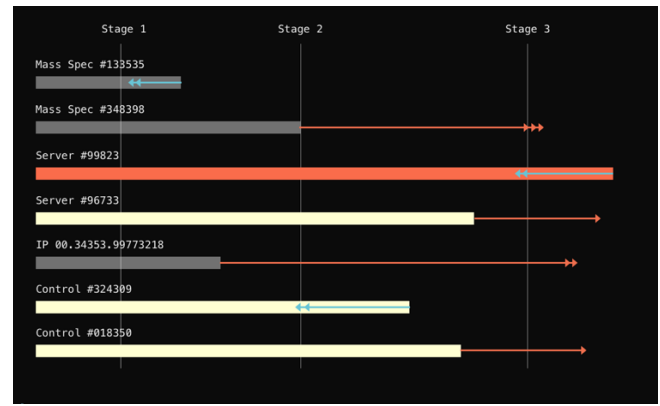

Fig. 3. CyPhyEye anomaly visualization concept based on the SEQUESTOR concept.

Each bar indicates the composite anomaly level of a single cyber-physical system, such as an instrument. Solid bars indicate the actual level of anomaly, while thinner lines represent the expected near future based on the observed rate of change. Arrows on the ends of the thin lines indicate the observed change velocity. Bar color indicates the raw level of anomaly concern for each system, and vertical lines represent the stage of action required.

A further visualization concept (Fig. 4) shows the relative amounts of message traffic that system components send and receive. The network visualization concept presents a physical map of the facility or area of operation with two circles for each communicating entity. The size of the solid circle represents the amount of sent traffic and the outline circle represents the amount of received traffic. The point-to-point connectors also show the relative amounts of traffic sent and received.

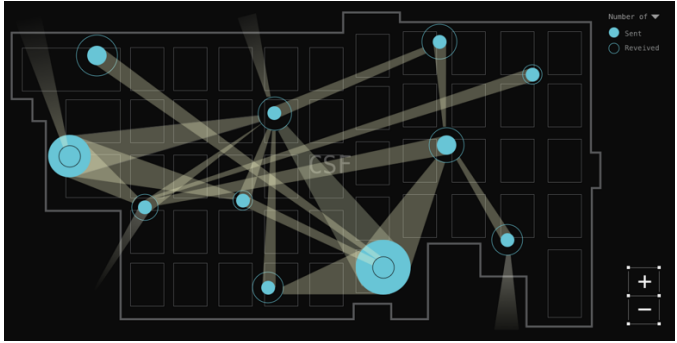

Fig. 4. CyPhyEye geographical communication visualization concept.

A final visualization concept for the user experience is the pathway visualization (Fig. 5). When a user selects a node on the communication visualization or a bar on the anomaly visualization, he or she may drill down to the pathway visualization, which shows the communication partners of a given cyber-physical system component and the relative amount of communication for each. Analysts may annotate the component with notes about its behavior, etc. This visualization will help both OT and IT analysts understand the communication patterns within the control system and isolate potential problems related to network traffic, errors, etc.

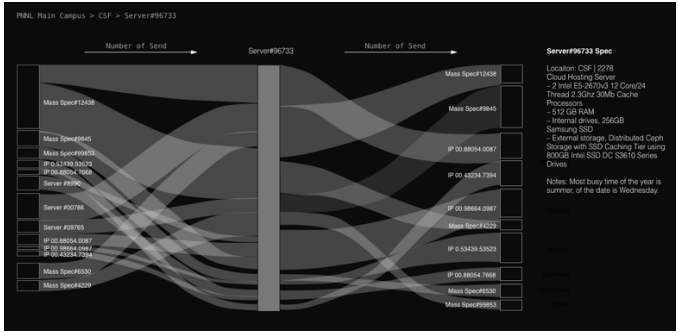

Fig. 5. Pathway visualization.

### B. Analytics Prototype

The anomaly metrics required by the visualization approach must be derived computationally in real time. We propose a deep-learning-based approach that can handle the massive streams of data from OT systems and alerts that cyber tools produce and rapidly identify what is normal and how anomalous event streams are.

Two kinds of communications data are available fro estimation of anomaly in OT systems: (1) communications within the field bus and (2) communications between field bus devices and external networks. Both sources provide an indication of the anomaly level of the OT systems. We plan to create three deep learning neural networks to monitor and control the OT system (Fig. 6).

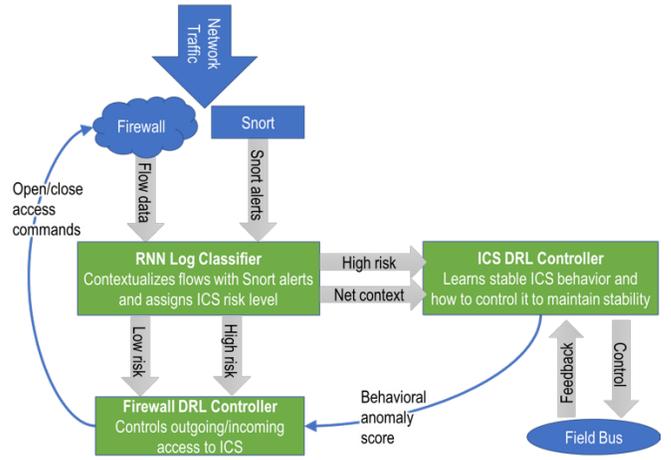

Fig. 6. Deep learning system to monitor anomaly levels in OT networks.

The first network will process network flow data and Snort alerts in the IT network, specifically looking for flows that contain OT protocols. This recurrent neural network (RNN) log classifier will be patterned on the work of [8], which makes heavy use of long short-term memory (LSTM) [9] deep neural network connected in series and analysis of textual log entries as both tokenized word vectors and character vectors. Their work is an unsupervised approach to detecting anomalous behavior in computer and network logs that is largely domain-agnostic. The model treats system logs as threads of interleaved "sentences" (event log lines) to train online unsupervised neural network language models using standard and bidirectional RNNs over network log data. They have created a tiered recurrent architecture, which provides context by modeling sequences of actions over time. The model has produced superior performance and fine-grained anomaly detection on open vocabulary logging sources. analytics engine was architected to compute anomaly metrics from field bus transactions, which are cyclic and repetitive, consisting of simple transactions. The network would be trained to classify network transactions as either low or high risk, initially based on their similarity to alerts from Snort, but later based on feedback from the other components. Additionally, this component produces a stream of network context data that contains flows that involve systems within the OT system being monitored.

The second component is planned to be a deep reinforcement learning (DRL) network whose job is to learn how to control the OT network in a manner similar to a BMS or NAC. This ICS DRL controller has an action set of enabling or disabling command/status information to/from the field bus, and its rewards come from producing the least anomalous behavior from the OT system. Essentially, this system consists of an autoencoder trained on stable field bus behaviors. The inputs to the model are the status messages on the field bus (the feedback), the high-risk behaviors detected by the first component, and the network context it provides. It produces control streams for the components of the field bus and a behavioral anomaly metric. The greater the distance metric between the feedback and the autoencoded results, the higher the behavioral anomaly metrics is produced.

The third component is planned to be another DRL network that takes the output of the RNN Log Classifier and the behavioral anomaly score from the ICS DRL controller and decides whether firewall rules should be adjusted to temporarily quarantine access between OT systems and the rest of the network. This approach is based on our prior work in [10]. The behavioral anomaly score forms the basis for the reward function for the DRL component. Additionally, over time, there will be an increasing reward for re-opening communication. We assume that allowed communication pathways are at least generally desirable, even if they sometimes produce undesirable results.

The behavioral anomaly metric, the raw flow and Snort data, and the classifications derived by the deep learning system will feed the visualization and enable humans to work together to maintain cyber safety in OT neworks.

## V. Conclusions and Future Work

Concern over cyber-attacks has led to the thoughtless proliferation of tools focused on addressing pressing cybersecurity needs in OT systems without long-term consideration of their utility. Much of what has been developed originates in the IT sector and has been inherited with little customization by the control systems world. As a result, control-systems professionals, who have not been traditionally responsible for security, now have a role in cyber security of computerized OT systems, and they lack tools customized to their environment. The data deluge presents difficulty while not solving the essential problem of lack of communication between IT and OT.

We have proposed a toolset that would perform much of the required analysis for the user and present only relevant information in a consistent way. The proposed situational awareness toolset will fuse information from ICS with cyber security data and present a common visualization. It will enable cross validation of automated warnings and enhance the quality of information that is presented to both IT and OT defenders through correlation of multiple data sources.

We plan to take these and other concepts back to the operators who helped us understand their needs to see how well the concept would work. Because the concept is driven by data that is available only in limited quantity in current systems, we plan to work with building owners and use our installed base of sensors to obtain the needed data to train our prototype.

Full definition of the architectures of the deep learning elements remains the largest area of future work. Given the data sources available and our access to both cyber security and OT operators we plan to use user-centric design methods to produce a usable prototype. Success of the prototype will be measured by how well defenders and operators are able to monitor and control their systems with and without our designs. We plan future papers to report on our findings.

## VI. Acknowledgments

Interviews and visualizations were funded by the Integrated Cyber-Physical Assessment and Response project of the U.S. Department of Energy's Integrated Joint Cybersecurity Coordination Center (iJC3) Cyber R&D Enterprise Cybersecurity Capability. This work was funded in part by the Deep Learning for Scientific Discovery investment at the Pacific Northwest National Laboratory. This paper corresponds to PNNL report PNNL-SA-138585.

[1] Alexander, LTG Keith, Address to the RSA Security Conference in San Francisco, CA, 21 April 2009. Transcript available at: https://www.nsa.gov/news-features/speeches-testimonies/speeches/21apr09-dir.shtml

[2] Department of Homeland Security Alert (IR-ALERT-H-16-056-01) Cyber-Attack Against Ukrainian Critical Infrastructure (2-25-16) (available at https://ics-cert.us-cert.gov/alerts/IR-ALERT-H-16-056-01 ). Accessed 25 Sept 2017.

[3] U.S. CERT (2018). VPNFilter Destructive Malware. Released 23 May 2018. https://www.us-cert.gov/ncas/current-activity/2018/05/23/VPNFilter-Destructive-Malware

[4] NSTAC, 2014. The President's National Security Telecommunications Advisory Committee (NSTAC) report to the President on the Internet of Things. Available at: http://www.dhs.gov/sites/default/files/publications/NSTAC%20Report%20to%20the%20President%20on%20the%20Internet%20of%20Things%20Nov%202014%20%28updat%20%20%20.pdf

[5] Department of Homeland Security (2014a), "Primary Stuxnet Advisory", ISC-CERT Advisory (ICSA-10-272-01), Original release: Sept. 29, 2010, revised: Jan. 21, 2014, available at: https://ics-cert.us-cert.gov/advisories/ICSA-10-272-01

[6] Krotofil, Marina, "Rocking the pocket book: Hacking chemical plants for competition and extortion." Black Hat White Paper, August 2015.

[7] Silvers, Robert, "Rethinking FISMA and Federal Information Security Policy," New York University Law Review, vol. 81, November, 2006 , pp. 1844–1874. Available from http://www.nyulawreview.org/sites/default/files/pdf/NYULawReview-81-5-Silvers.pdf .

[8] Tuor, Aaron, Ryan Baerwolf, Nicolas Knowles, Brian Hutchinson, Nicole Nichols, Rob Jasper (2018). "Recurrent Neural Network Language Models for Open Vocabulary Event-Level Cyber Anomaly Detection" in Proceedings of Artificial Intelligence in Cyber Security Workshop; AAAI-2018

[9] Hochreiter, Sepp and Jürgen Schmidhuber (1997). "Long short-term memory". Neural Computation. 9 (8): 1735–1780. PMID 9377276.

[10] Oehmen CS, Carroll TE, Paulson PC, Best DM, Noonan CF, Thompson SR, Jensen JL, Fink GA, and Peterson ES, 2015. "Behavior-dependent Routing: Responding to Anomalies with Automated Low-cost Measures." In *Proceedings of the 2015 Workshop on Automated Decision Making for Active Cyber Defense* (SafeConfig '15). ACM, New York, NY, USA, 55-58.